\documentclass[aps,prl,twocolumn,superscriptaddress]{revtex4}

\usepackage{notes2bib}
\usepackage{graphicx}
\usepackage{dcolumn}
\usepackage{bm}
\usepackage{float}
\usepackage{subfigure}
\usepackage{amsmath} 
\usepackage{amsthm} 
\usepackage{amssymb}	
 
\let\baraccent=\= 
\renewcommand{\=}[1]{\stackrel{#1}{=}} 

\theoremstyle{definition}

\theoremstyle{remark}

\newcommand{\om}{\omega}

\usepackage{amssymb,amsfonts,amsmath}

\begin{document}


\title{Noise-induced quantum coherence drives photocarrier generation dynamics at polymeric semiconductor heterojunctions}

\author{Eric R.\ Bittner}
\address{ Department of Chemistry and Physics, University of Houston, Houston, TX 77204}
\address{Department of Physics and
Regroupement qu{\'e}b{\'e}cois sur les mat{\'e}riaux de pointe, Universit{\'e} de Montr{\'e}al,
C.P.\ 6128, Succursale centre-ville, 
Montr{\'e}al (Qu{\'e}bec) H3C 3J7, Canada}

\author{Carlos Silva}
\address{Department of Physics and
Regroupement qu{\'e}b{\'e}cois sur les mat{\'e}riaux de pointe, Universit{\'e} de Montr{\'e}al,
C.P.\ 6128, Succursale centre-ville, 
Montr{\'e}al (Qu{\'e}bec) H3C 3J7, Canada}

\begin{abstract}
We present a fully quantum-mechanical model of the electronic dynamics of primary photoexcitations in a 
polymeric semiconductor heterojunction, 
which includes both polymer stacking and phonon relaxation. 
By examining the phonon-induced fluctuations in the state-to-state energy gaps and the 
exact golden-rule rate constants, we conclude that resonant tunnelling 
between the primary exciton to delocalized interchain charge-transfer states may be the 
initial step in the formation of charge-carriers in polymer-based bulk-heterojunction photovoltaic diodes.
\end{abstract}


\maketitle



\section{Introduction}

Photovoltaic diodes based on blends of semiconductor polymers and fullerene derivatives now produce power conversion efficiencies exceeding 10\% under standard solar illumination~\cite{He:2012uq}, demonstrating that photocarriers can be generated efficiently in well optimized organic heterostructures. 
Recent advances in ultrafast spectroscopic techniques have advanced our understanding of quantum dynamics to the point where quantum coherences between the relevant states can be observed and interpreted as playing a crucial role in the efficiency of photosynthesis in biological systems~\cite{1367-2630-12-6-065042,Ishizaki:2009gd,doi:10.1021/jz900062f,yang:045203,Harel17012012} and semiconductor polymers~\cite{Collini16012009,scholes2003,Beljonne:2005}. 
Numerous ultrafast spectroscopic measurements have reported that charged photoexcitations in these systems can be generated on $\leq 100$-fs timescales~\cite{Sariciftci:1994kx,Banerji:2010vn,Jailaubekov:2013fk,Grancini:2013uq}, but full charge separation to produce photocarriers is expected to be energetically expensive given strong Coulombic barriers due to the low dielectric constant in molecular semiconductors. Nonetheless, G\'elinas {\em et~al.}\ have put forth that electrons and holes separate by 4\,nm over the first 100\,fs~\cite{Gelinas:2013fk}, and evolve further on picosecond timescales to produce unbound charge pairs. Such spectacularly rapid charge separation points strongly to quantum coherence dynamics which are correlated to the 
dynamical motion of the molecular framework~\cite{Rozzi:2013fk}.  The significant element in the context of ultrafast charge separation in the system considered 
here is the involvement of delocalized charge-transfer states in the early quantum dynamics of the exciton. 
A detailed mechanistic understanding of primary charge generation dynamics is of key fundamental importance in the development of organic solar cells.


\begin{figure}[t]
\includegraphics[width=0.6\columnwidth]{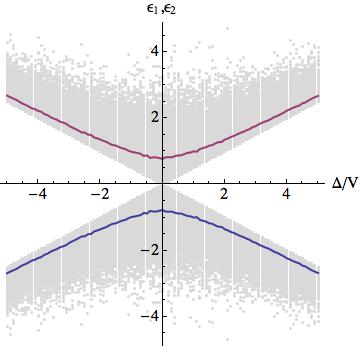}
\caption{Energy eigenvalues distribution of equation (1) vs. energy gap $\Delta$ over multiple realizations of the coupling 
with $W = 1$. 
The  solid curves are $\pm \sqrt{\Delta^{2}+W^{2}}/2$ and dashed lines  are $ \pm\Delta/2$.}\label{figure1}
\end{figure}

By quantum coherence we specifically mean the \emph{temporal} phase coherence of a single quantum mechanical system between possible asymptotic states
that arises due to the quantum mechanical time-evolution of the system's wave function. 
  Coherence loss (decoherence) occurs when a single quantum  system is in contact 
with thermal or noisy environment or when an external measurement is performed on the system such in stimulated emission
\bibnote{Thus, within the framework of this paper, the outcome of the coherence dynamics does not depend on whether the excitation is impulsive, as in ultrafast spectroscopies, or continuous, as in solar illumination.}.

\begin{figure*}[t]
\subfigure[]{\includegraphics[width=\columnwidth]{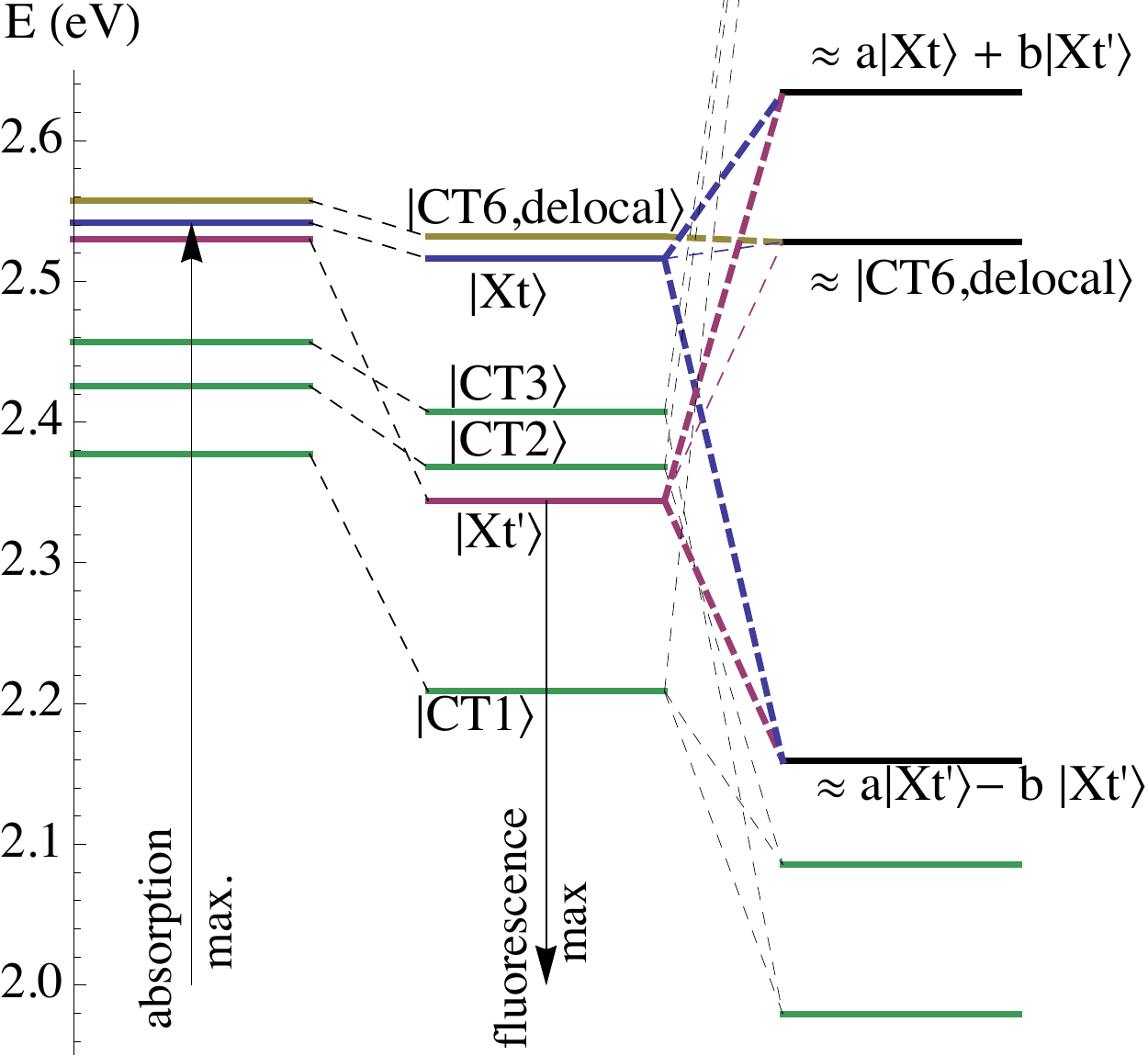}}
\subfigure[]{\includegraphics[width=\columnwidth]{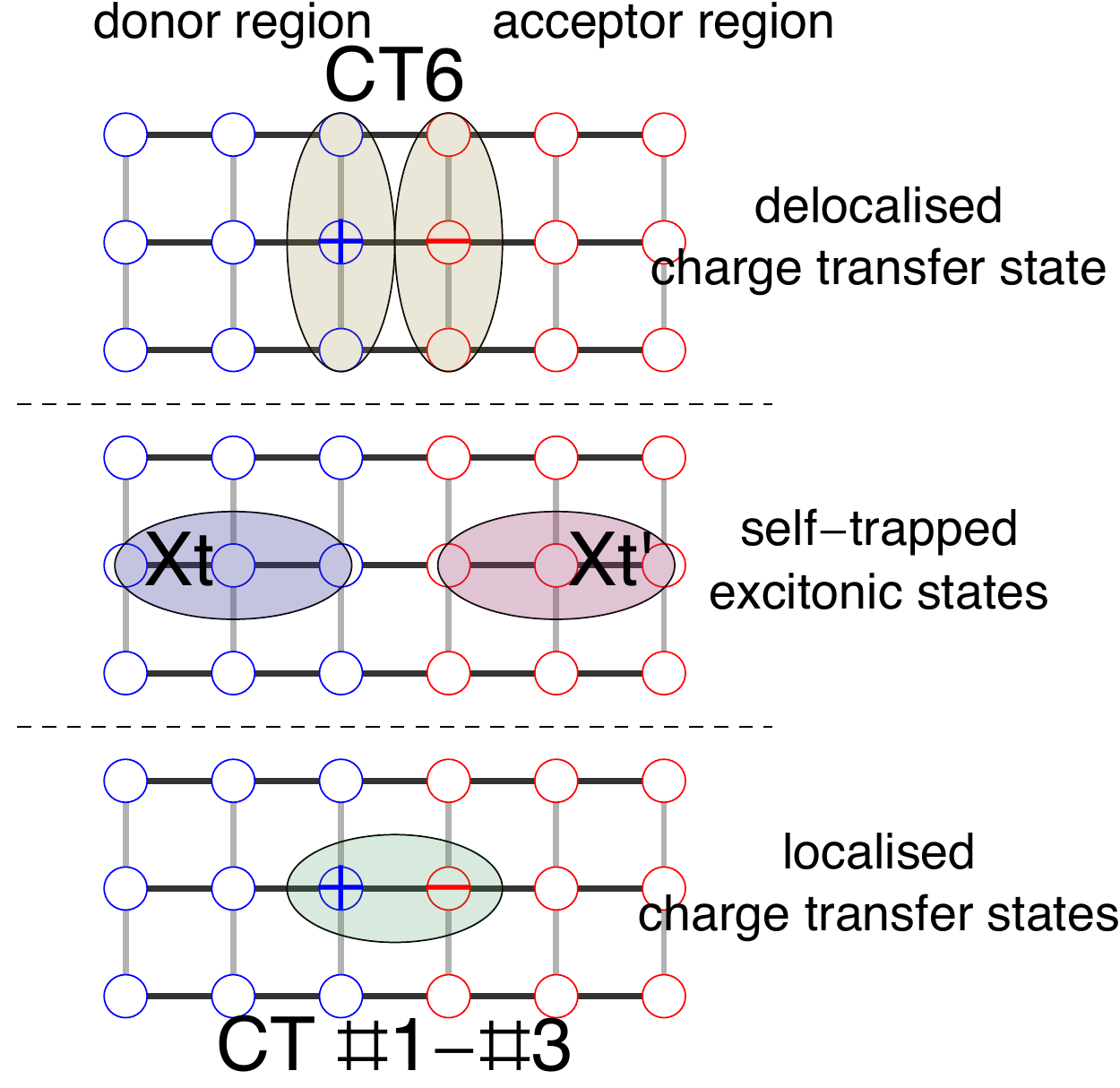}}
\caption{(a) Vertical, adiabatic/relaxed, and fluctuation-mixed energy levels for the model heterojunction system. 
Above state \#6, the density of states is essentially continuous. 
Arrows indicate the states with the most oscillator strength to the ground-state  (b) Cartoon sketch of the types of low-lying 
states generated by our model.
Color coding of the energy levels corresponds to the wave functions depicted in (b).
Dashed lines between levels indicate ``parentage'' and relative mixing.
 }\label{figure2}
\end{figure*}


%


\begin{figure*}[h]
%
\subfigure[]{\includegraphics[width=0.3\columnwidth]{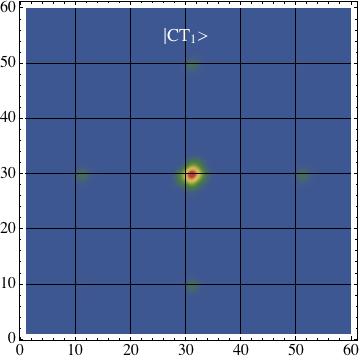}}
\subfigure[]{\includegraphics[width=0.3\columnwidth]{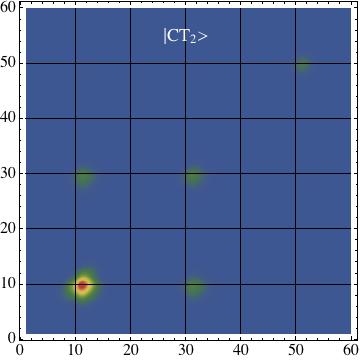}}
\subfigure[]{\includegraphics[width=0.3\columnwidth]{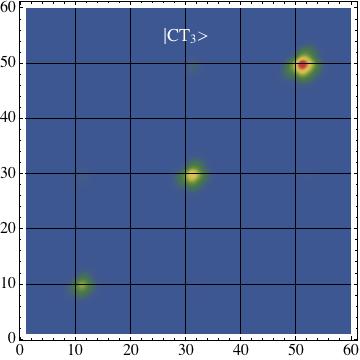}}

\subfigure[]{\includegraphics[width=0.3\columnwidth]{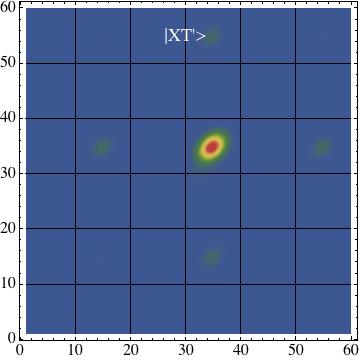}}
\subfigure[]{\includegraphics[width=0.3\columnwidth]{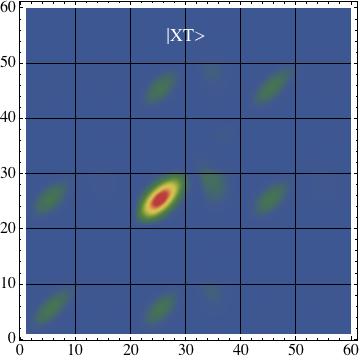}}
\subfigure[]{\includegraphics[width=0.3\columnwidth]{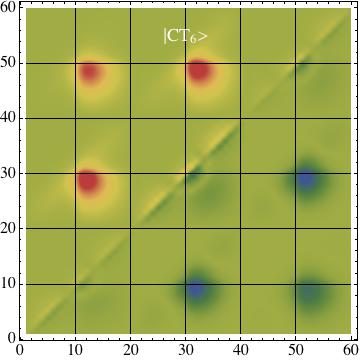}}
\caption{Relaxed energy eigenstates of the 0.5eV off-set model in a site-wise representation. The horizontal axis denotes the location of the valance-band
hole and the vertical axis denotes the location of the conduction-band electron.  
The grid-lines denote polymer segments.  Polymer 1 spans 
sites 1--20, polymer 2 spans sites 21--40, and polymer 3 spans sites 41--60. }\label{figure3}
\end{figure*}

We here present  a simple model followed by rigorous  quantum dynamical calculations that underscore the 
role that quantum coherence and energy fluctuations may play in the production of charge-separated states in 
polymeric type-II heterojunction devices.   Similar noise-induced coherent processes may also enhance the quantum efficiency in light-harvesting complexes~\cite{Dorfman19022013,Chin:2013fk} and be exploited to reduce radiative 
recombination in photovoltaics cells~\cite{PhysRevLett.104.207701}.
We show that resonant tunnelling processes brought about 
by environmental fluctuations produce delocalized charge-transfer states on $<100$-fs timescales, which may readily produce photocarriers. 
\section{Noise induced coherence}
A simple model for this can be developed as follows. 
Consider two electronic levels separated by an  energy gap $\Delta$ 
coupled to a noisy phonon environment that modulates their instantaneous coupling 
$dV(t)$. The dynamics of the coupled system are described by 
\begin{eqnarray} 
H = -\frac{1}{2}\Delta \sigma_{z} + \frac{1}{2}dV(t)\sigma_{y},
\end{eqnarray} 
where $\bar{dV} = 0$ and the fluctuations in $dV(t)$  introduce uncertainty in the energy of the system 
\begin{eqnarray} 
 W^{2} = \overline{dV^{2}}  = \int_{-\infty}^{+\infty}\frac{d\omega}{2\pi} S(\omega),
\end{eqnarray} 
where $S(\omega)$ is the spectral density of the environment. 
Fig.~1 shows the energy eigenvalues of Eq.~1 for multiple samplings of the 
off-diagonal coupling $dV(t)$ with fixed $W = 1$.  
The superimposed red and blue curves give the average 
energy eigenvalue at each $\Delta$.  These correspond to the 
eigenvalues of the noise-averaged Hamiltonian.
\begin{eqnarray} 
\langle H \rangle= -\frac{1}{2}\Delta \sigma_{z} + \frac{1}{2}W\sigma_{y}.
\end{eqnarray}

If we consider the time-evolution of a state under the noisy conditions,
we can identify two regimes.  One is 
where the fluctuations are greater than the gap, $W \gg \Delta$.
 This is the resonant tunnelling  regime where the states are 
 strongly mixed by the coupling. 
  Secondly, when $W\ll \Delta$,  the 
coupling  is  perturbative and induces coherent oscillations between the two  eigenstates of $\sigma_{z}$.
Factoring out a common phase factor, we can write the time-evolved state as
\begin{eqnarray}
|\psi(t) \rangle =(\sqrt{1-a^{2}} |0\rangle  + e^{i\Omega t}a |1\rangle),
\end{eqnarray}
where $\Omega =  (\epsilon_{1} - \epsilon_{0})/\hbar = \sqrt{\Delta^{2} + V^{2}}/\hbar$ is the relative phase
(i.e. the Rabi frequency) between the basis states.
 Taking $W \ll \Delta$, we have an average Rabi frequency of 
$
\hbar\bar\Omega \approx \Delta + {W^{2}}/{2\Delta} $.  Since the second term 
in this expansion originates from the noise, we can define a  decoherence rate of 
$
T_{d}^{-1} = {W^{2}}/2 \hbar \Delta
$
which is essentially the Kubo formula for the dephasing rate~\cite{kubo:174,Kubo:1962fk}. 
In the resonant tunnelling regime, on the other hand, when the 
gap is small compared to the fluctuations,  both the average Rabi frequency and 
dephasing time are proportional to the fluctuations, 
$
\hbar\Omega \approx W + {\Delta^{2}}/{2W} $.  Taking $\Delta \to 0$ and averaging over noise gives
a decoherence rate 
$T_{d}^{-1}  = W/\hbar$.

The role of coherence becomes clear when we consider the population transition rates between 
states.  We can easily write the equations of motion for both the
coherences and populations and show 
that they reduce to that of a damped oscillator
with a decay constant given by
\begin{eqnarray}
k = \frac{\bar\Omega^{2}}{2} \frac{T_d}{1+(T_d\Delta /\hbar)^2}.\label{gr-approx}
\end{eqnarray}
As $T_{d}\to 0$, the transition rate vanishes. This is a manifestation of  
the quantum Zeno effect where rapid quantum measurements on the 
system collapses any superposition states that form due to the interactions.
In the resonant tunnelling regime considered here, where the  coupling 
between states is due to the phonon fluctuations, the fluctuations  generate 
coherences and give rise to damped Rabi oscillations between otherwise uncoupled states. 

Such coherences may be exploited in the case where an initial photoexcitation (exciton)
is in the resonant tunnelling regime with states that are not 
 photoexcited directly, such as charge-transfer states.   Such non-thermalized ``hot-exciton'' states 
have been implicated recently as important precursors in the generation of photocurrent in organic photovoltaics
since they undergo fission within the first 50\,fs following excitation,
creating both interfacial charge transfer states (CTSs) and polaron species in low-bandgap polymer system~\cite{Jailaubekov:2013fk,Grancini:2013uq}.

\section{Heterojuction Lattice Model}
To explore the role of coherence in such systems, 
we constructed a fully quantum-mechanical/finite temperature model for a polymer heterojunction  
consisting of three parallel stacked polymers each with 20 sites, with an energy off-set  between donor and an acceptor regions of $\Delta E = 0.5$\,V.
Each site contributes a valance and a conduction band Wannier orbital and the electronic ground state is where
 each site is doubly occupied.    We estimate the interchain hopping term to be an order of magnitude 
smaller than the intrachain hopping term, $t_{\perp} \approx t_{\parallel}/10$, such that electron or hole
motion along the chains is easier than hopping between chains.   
 Single electron/hole 
excitations from the ground-state are considered within 
configuration interaction (CI) theory. 
Each site also contributes two localized  phonon modes which modulate the local 
energy gap at each site. Linear coupling between localized phonons give rise to 
optical phonon bands that are delocalized over each polymer chain. 
The electron/phonon couplings were determined by  experimental Huang-Rhys factors for
poly-pheneylene-vinylene type polymers.

We have used this approach previously to describe the energetics,  dynamics, and spectroscopy of polymer-based
donor-acceptor systems and diodes~\cite{karabunarliev:057402,Bittner:2005be,karabunarliev:3988}.  
Moreover, the  model can provide
the necessary input for describing the dynamics of a polymer-based photovoltaic cell~\cite{pereverzev:104906,tamura:021103,tamura:107402}.
A similar lattice model for a bulk-heterojunction 
was  presented recently by Troisi that includes many of the features of our model, 
but does not include explicit phonons and electronic transitions are introduced via the semiclassical
Marcus theory~\cite{Troisi2013}.   
Further details and parameters of our model are included as supplementary material.
The energetics and time-scales produced by our model are consistent   with photon-echo experiments on 
MeH-PPV systems \cite{PhysRevB.71.045203} and 
with the fully atomistic 
quantum/classical 
molecular dynamics simulations on  phthalocyanine/fullerene interfaces
 reported in Ref.~\cite{Jailaubekov:2013fk}.

\begin{figure}[t]
\centering
\includegraphics[width=0.64\columnwidth]{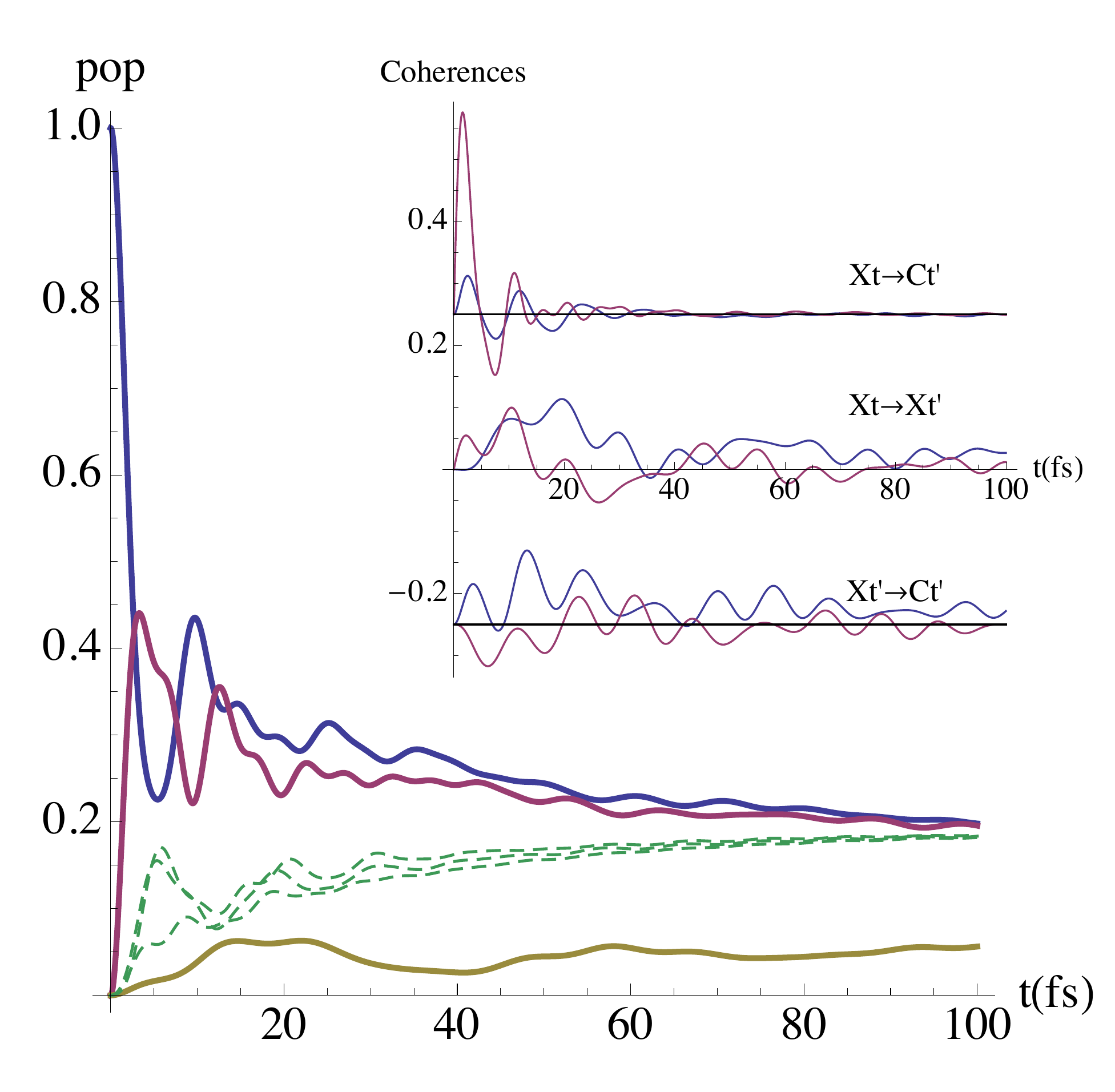}
\caption{Noise-averaged population and coherence (inset) dynamics for the model heterojunction system.
Both the real and imaginary components  of the coherences are shown.
 The population curves in (a) correspond to the primary exciton $Xt$ (blue), the secondary exciton $Xt'$ (red),  delocalized CT state (gold), 
and localized or interface-pinned CT states (green).}\label{figure4}
\end{figure}

Fig.~\ref{figure2} shows the six lowest energy levels of the $60^2$ CI eigenstates 
produced by our model and a sketch of the relevant states. The full density of states and spectroscopic 
properties of our model are given in the supplementary material. 
In  Fig.~\ref{figure3} we show the relaxed transition densities in terms of the electron/hole configurations
 for the  six lowest energy states of our model system with an energy 
off-set of 0.5\,eV.  The configurations are indexed such that polymer 1 spans sites 1--20, polymer 2 spans sites 21--40, and polymer 3 spans sites 41--60 with the energy off-set between 
sites 10 and 11, sites 30 and 31, and sites 50 and 51 respectively. 
States \#1--\#3 correspond charge-separated states that are pinned to the interface with some degree of delocalisation 
between the parallel chains. 
We denote these as $|CT_{n}\rangle$. 

State \#4 and \#5 both are largely excitonic in character and correspond to excitons localized on 
side of the heterojunction or the other. 
These we denote as $|XT'\rangle$ and $|XT\rangle$, respectively.
The parent (unrelaxed) state of \#5 carries the most oscillator strength to the ground state of the system and
we consider it to be our primary exciton.  It relaxes and localizes to the ``donor'' side of the system. 
The parent state of \#4 is also excitonic, but is delocalized over both the donor and acceptor sides with a 
node at the band-offset. Because of the node, it carries little oscillator strength. 
However, upon relaxation, it localizes to the ``acceptor''  side and carries most of the 
oscillator strength for emission to the ground-state.
These segment-localized exciton states are equivalent to the ``Localized Exciton Ground States'' (LEGS) discussed 
recently by Barford {\em et al.}~\cite{doi:10.1021/jp307040d,doi:10.1021/jp307041n}
and are likely responsible for the rapid excitonic transfer along a polymer chain~\cite{doi:10.1021/jz900062f}.

\begin{figure*}[H]
\centering
\includegraphics[width=0.64\columnwidth]{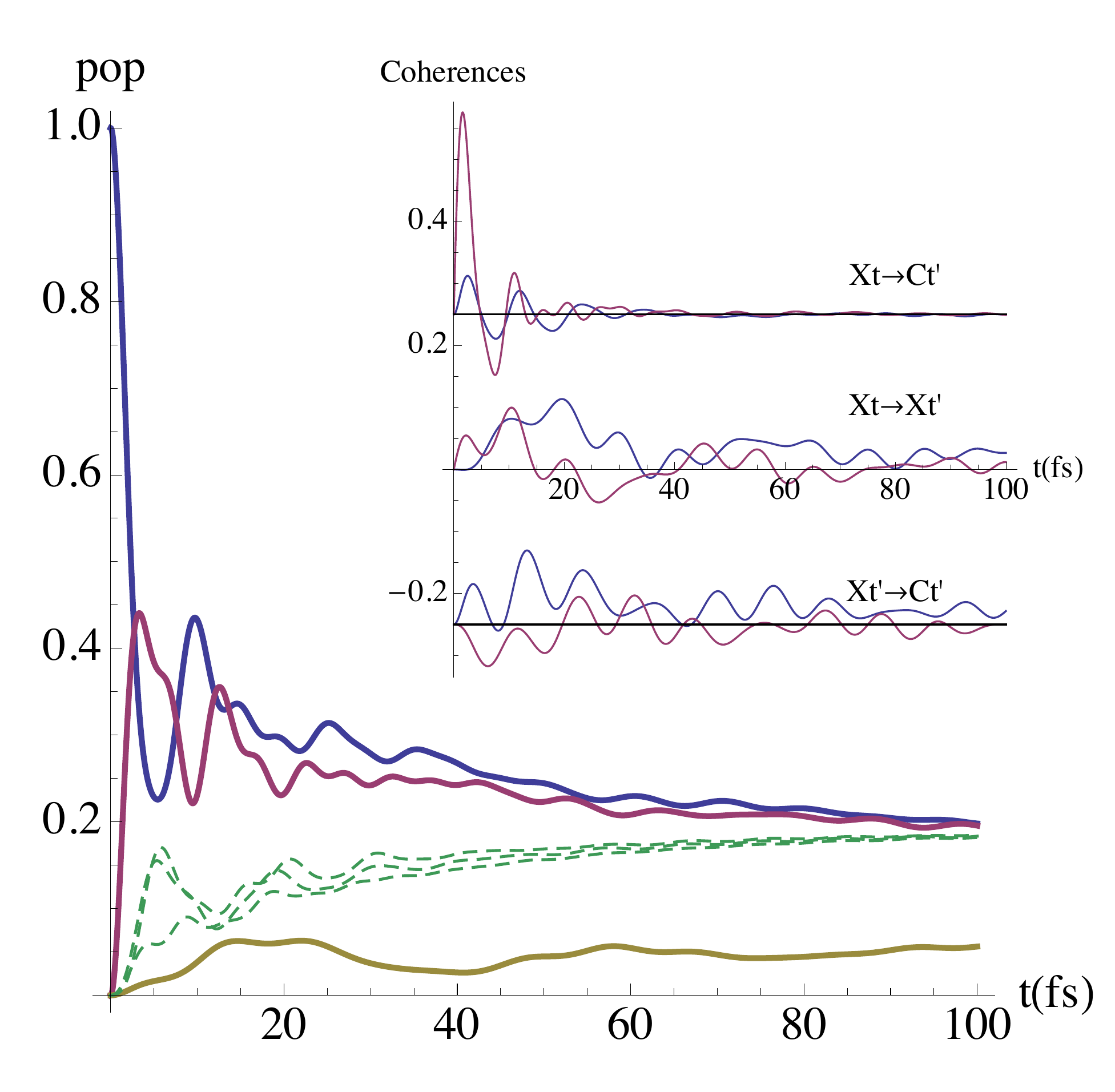}
\caption{Noise-averaged population and coherence (inset) dynamics for the model heterojunction system.
Both the real and imaginary components  of the coherences are shown.
 The population curves in (a) correspond to the primary exciton $Xt$ (blue), the secondary exciton $Xt'$ (red),  delocalized CT state (gold), 
and localized or interface-pinned CT states (green).}\label{figure4}
\end{figure*}

Interestingly, state \#6 is a charge-transfer state that is delocalized over all three chains.  It is also
slightly higher in energy than the $|XT\rangle$ state.  Such states could serve as precursor  states for the 
rapid formation of independent 
polaron states with the electron and hole localized on different chains and are likely the
 primary source for photocurrent in bulk 
heterojunction systems~\cite{bittner:034707,Troisi2013}. 
Above state \#6 is largely a continuum of electron/hole eigenstates that carry little or no oscillator strength. 
We focus our attention on states near the primary exciton states.

\subsection{Phonon noise}
Central to our model is the notion that off-diagonal couplings are due to fluctuations in the phonon degrees of freedom.
For this we perform a polaron transform on our model Hamiltonian using the unitary transformation
\begin{eqnarray}
U=e^{-\sum_{ni}\!\!\frac{g_{nni}}{\hbar\om_i}|n\rangle \langle
n|(a^{\dagger}_i-a_i)}= \sum_{n}|n\rangle \langle
n|e^{-\sum_{i}\!\!\frac{g_{nni}}{\hbar\om_i}(a^{\dagger}_i-a_i)}
\label{unitary} 
\end{eqnarray} 
under which our transformed Hamiltonian can be written in terms of the 
diagonal elements 
\begin{eqnarray} \tilde H_0=U^{-1}H_0U
=\sum_n\tilde\epsilon_n |n\rangle \langle n|+\sum_i\om_ia^{\dagger}_ia_i,
 \end{eqnarray}
where the renormalized electronic energies are 
\begin{eqnarray}
\tilde\epsilon_n=\epsilon_n-\sum_{i}\frac{g_{nni}^2}{\hbar\omega_i},  
\end{eqnarray}
and off-diagonal terms
\begin{eqnarray} \hat V_{nmi}=g_{nmi}\left(a^{\dagger}_i+
a_i-\frac{2g_{nni}}{\hbar\omega_i}\right)e^{\sum_{j}\frac{(g_{nnj}-g_{mmj})}{\hbar\om_j}(a^{\dagger}_j-a_j)}.
\label{opm}
\end{eqnarray}
In the transformed picture the
electronic transitions from state $|n\rangle$ to $|m\rangle$ are dressed by the 
excitations  of all the normal phonon modes.  
Transforming to the interaction representation and performing a trace over the 
phonons gives the spectral density in terms of the autocorrelation of the electron/phonon coupling operators.
\begin{eqnarray}
S_{nm}(\omega) = \int_{-\infty}^{\infty} dt e^{-i\omega t}\langle \hat V_{nm}(t) \hat V_{mn}(0)\rangle.\label{gr-exact}
\end{eqnarray}
Note that the Golden rule transition rates can be computed by evaluating $S_{nm}(\omega)$ at the transition frequency, 
$\omega_{nm} = (\tilde \epsilon_{n}-\tilde \epsilon_{m})/\hbar$.
The derivation and explicit form for these in terms of the 
non-adiabatic electron/phonon couplings is quite lengthy and is given in Ref.~\cite{pereverzev:104906}. 
Integrating over phonon frequency gives the r.m.s.\  energy fluctuations
and the average  coupling strength between each pair of states.  
Table 1 in the Supplementary material summarizes the energetics, fluctuations, and 
computed transition rates between various states in our model.

The right-most set of energy levels in Fig.~2(a) indicate the mixing between the adiabatic/relaxed 
electronic states due to the fluctuations in the off-diagonal (i.e.\ non-adiabatic) coupling.  
At 10\,K, the fluctuations are entirely due to zero-point motion in the phonon degrees of freedom. 
It is also important to point out that these phonon-mixed states are not stationary eigenstates of the 
system.  That distinction belongs to the left-most and middle set of states
which are stationary eigenstates of the electronic Hamiltonian 
and correspond to the spectroscopic observables.  The phonon-mixed 
states on the right are superpositions of the adiabatic states and the coherence brought about by the 
mixing also decays rapidly giving rise to the 
homogeneous line-broadening in the spectroscopic transitions.

\subsection{Noise-averaged Dynamics}

In Fig. 4 we consider the population and coherence dynamics for the noise-averaged system.
 First the $Xt \to Xt^{\prime}$ transfer, which corresponds to the 
transfer of a donor-side exciton to a self-trapped exciton on the
acceptor side, occurs very rapidly forming a superposition within the first 5--10 fs. 
The golden-rule rate for this steps gives a time-scale of 17\,fs with a 2.5-fs coherence time. 
The rapid loss of coherence can be understood 
from the fact that energy difference between $Xt$ and $Xt^{\prime}$ is largely due to the 
reorganization of the phonon lattice about $Xt^{\prime}$ and hence strong electron/phonon coupling leads to both rapid 
decoherence and mixing between these two states.
This transition appears to be 
the gateway for subsequent relaxation into the lower lying charge-transfer states. 
Interestingly, in spite of the short coherence time between the $Xt$ and $Xt'$ states, 
coherences persist for over  100\,fs.  This long-lived coherence is due to the indirect coupling  of the 
secondary $Xt'$ state to the delocalized $CT6$ state.

The second process that stands out is the {\em upward} transition from the primary exciton 
to the  delocalized $CT6$ state.   This state is only 
16\,meV in energy above the primary exciton.   Since $W/|\Delta| \approx 2$, the fluctuations are
 larger than the gap itself,  and these states are within the resonant tunnelling regime; however, with 
 coherence time of 24\,fs. Moreover, the noise-induced splitting between the two excitonic states
 places the delocalized $CT6$ state within the homogeneous line shape of the primary exciton thereby 
promoting rapid dissociation of the primary exciton into these weakly-bound charge-separated states.
Our golden-rule estimate for the $Xt$ to delocalized $CT$ transition time is 271\,fs with a coherence timescale of 24\, fs.
{\em This strongly suggests that the 
initial decay of an exciton to form separated polaron pairs may 
occur via resonant tunnelling processes and that this is the initial step in the formation 
of charge carriers in a bulk-heterojunction system. }

Recent ultrafast measurement of charge-transfer-exciton \emph{population} dynamics suggest that current producing states are generated 
on the $\approx 100$\,fs timescales in systems as diverse as model heterojunctions between Cu-pthalocyanine donors and C$_{60}$~\cite{Jailaubekov:2013fk}, and bulk heterojunction blends of push-pull conjugated polymers~\cite{Grancini:2013uq} and molecules~\cite{Gelinas:2013fk} with fullerene derivatives, and that excitation with excess energy above the optical gap enhances the yield of photocarriers. We propose that the \emph{coherence} dynamics drive this process as described in this communication.  Direct probes of these dynamics are accessible by contemporary ultrafast coherent spectroscopies involving phase-locked femtosecond pulse sequences,  thus the proposed role of quantum coherence can be tested in state-of-the-art  
heterostructures implemented in polymer-based photovoltaic diodes.

\section{Summary}
In summary, we have examined the role of quantum coherence  in charge separation at a polymer type-II heterojunction by implementing a model that takes into account two-dimensional electronic dispersion in polymer stacks and phonons.  Our lattice model is generic and as such the physics and dynamics produced by 
the model should be ubiquitous 
over a wide range of organic bulk-heterojunction systems.
We conclude that resonant tunnelling between excitons to delocalized interchain charge-transfer states during the course of decoherence may be the initial step in the formation of photocarriers.

\begin{acknowledgments}
We thank Simon G\'elinas and Sir Richard Friend for sharing ref.~\cite{Gelinas:2013fk} prior to publication. 
The work at the University of Houston was funded in part by the National Science Foundation (CHE-1011894) and the Robert A. Welch Foundation (E-1334).
CS acknowledges support from the Canada Research Chair in Organic Semiconductor Materials.
ERB acknowledges support from  Fulbright Canada and the US Department of State.  The authors declare no  competing interests.  Both authors
participated in the conception and writing of this paper. The calculations were performed by ERB.
\end{acknowledgments}

%

\begin{thebibliography}{10}

\bibitem{He:2012uq}
Zhicai He, Chengmei Zhong, Shijian Su, Miao Xu, Hongbin Wu, and Yong Cao.
\newblock {Enhanced power-conversion efficiency in polymer solar cells using an
  inverted device structure}.
\newblock {\em {Nature Photonics}}, {6}({9}):{591--595}, {SEP} {2012}.

\bibitem{1367-2630-12-6-065042}
Dugan Hayes, Gitt Panitchayangkoon, Kelly~A Fransted, Justin~R Caram, Jianzhong
  Wen, Karl~F Freed, and Gregory~S Engel.
\newblock Dynamics of electronic dephasing in the fenna--matthews--olson
  complex.
\newblock {\em New Journal of Physics}, 12(6):065042, 2010.

\bibitem{Ishizaki:2009gd}
Akihito Ishizaki and Graham~R. Fleming.
\newblock Theoretical examination of quantum coherence in a photosynthetic
  system at physiological temperature.
\newblock {\em Proceedings of the National Academy of Sciences},
  106(41):17255--17260, 10 2009.

\bibitem{doi:10.1021/jz900062f}
Gregory~D. Scholes.
\newblock Quantum-coherent electronic energy transfer: Did nature think of it
  first?
\newblock {\em The Journal of Physical Chemistry Letters}, 1(1):2--8, 2010.

\bibitem{yang:045203}
Xiujuan Yang, Tieneke~E. Dykstra, and Gregory~D. Scholes.
\newblock Photon-echo studies of collective absorption and dynamic localization
  of excitation in conjugated polymers and oligomers.
\newblock {\em Physical Review B (Condensed Matter and Materials Physics)},
  71(4):045203, 2005.

\bibitem{Harel17012012}
Elad Harel and Gregory~S. Engel.
\newblock Quantum coherence spectroscopy reveals complex dynamics in bacterial
  light-harvesting complex 2 (lh2).
\newblock {\em Proceedings of the National Academy of Sciences},
  109(3):706--711, 2012.

\bibitem{Collini16012009}
Elisabetta Collini and Gregory~D. Scholes.
\newblock Coherent intrachain energy migration in a conjugated polymer at room
  temperature.
\newblock {\em Science}, 323(5912):369--373, 2009.

\bibitem{scholes2003}
Gregory~D. Scholes.
\newblock Long-range resonance energy transfer in molecular systems.
\newblock {\em Annual Review of Physical Chemistry}, 54:57, 2003.

\bibitem{Beljonne:2005}
D~Beljonne, E~Hennebicq, C~Daniel, L~M Herz, C~Silva, G~D Scholes, F~J~M
  Hoeben, P~Jonkheijm, A~P H~J Schenning, S~C~J Meskers, R~T Phillips, R~H
  Friend, and E~W Meijer.
\newblock Excitation migration along oligophenylenevinylene-based chiral
  stacks: delocalization effects on transport dynamics.
\newblock {\em J Phys Chem B}, 109(21):10594--10604, 2005.

\bibitem{Sariciftci:1994kx}
NS~Sariciftci and AJ~Heeger.
\newblock {Reversible, Metastable, Ultrafast Photoinduced Electron-Transfer
  from Semiconducting Polymers to Buckminsterfullerene and in the Corresponding
  Donor-Acceptor Heterojunctions}.
\newblock {\em {International Journal of Modern Physics B}},
  {8}({3}):{237--274}, {JAN 30} {1994}.

\bibitem{Banerji:2010vn}
Natalie Banerji, Sarah Cowan, Mario Leclerc, Eric Vauthey, and Alan~J. Heeger.
\newblock {Exciton Formation, Relaxation, and Decay in PCDTBT}.
\newblock {\em {Journal of the American Chemical Society}},
  {132}({49}):{17459--17470}, {DEC 15} {2010}.

\bibitem{Jailaubekov:2013fk}
Askat~E. Jailaubekov, Adam~P. Willard, John~R. Tritsch, Wai-Lun Chan, Na~Sai,
  Raluca Gearba, Loren~G. Kaake, Kenrick~J. Williams, Kevin Leung, Peter~J.
  Rossky, and X-Y. Zhu.
\newblock Hot charge-transfer excitons set the time limit for charge separation
  at donor/acceptor interfaces in organic photovoltaics.
\newblock {\em Nat Mater}, 12(1):66--73, 01 2013.

\bibitem{Grancini:2013uq}
G.~Grancini, M.~Maiuri, D.~Fazzi, A.~Petrozza, H-J. Egelhaaf, D.~Brida,
  G.~Cerullo, and G.~Lanzani.
\newblock Hot exciton dissociation in polymer solar cells.
\newblock {\em Nat Mater}, 12(1):29--33, 01 2013.

\bibitem{Gelinas:2013fk}
Simon G{\'e}linas, Akshay Rao, Abhishek Kumar, Samuel~L. Smith, Alex~W. Chin,
  Jenny Clark, Thomas~S. van~der Poll, Guillermo~C. Bazan, and Richard~H.\
  Friend.
\newblock Coherent charge separation in organic semiconductor photovoltaic
  diodes.
\newblock {\em Science}, Submitted.

\bibitem{Rozzi:2013fk}
Carlo Andrea~Rozzi, Sarah Maria~Falke, Nicola Spallanzani, Angel Rubio, Elisa
  Molinari, Daniele Brida, Margherita Maiuri, Giulio Cerullo, Heiko Schramm,
  Jens Christoffers, and Christoph Lienau.
\newblock Quantum coherence controls the charge separation in a prototypical
  artificial light-harvesting system.
\newblock {\em Nat Commun}, 4:1602, 03 2013.

\bibitem{Note1}
Thus, within the framework of this paper, the outcome of the coherence dynamics
  does not depend on whether the excitation is impulsive, as in ultrafast
  spectroscopies, or continuous, as in solar illumination.

\bibitem{Dorfman19022013}
Konstantin~E. Dorfman, Dmitri~V. Voronine, Shaul Mukamel, and Marlan~O. Scully.
\newblock Photosynthetic reaction center as a quantum heat engine.
\newblock {\em Proceedings of the National Academy of Sciences},
  110(8):2746--2751, 2013.

\bibitem{Chin:2013fk}
A.~W. Chin, J.~Prior, R.~Rosenbach, F.~Caycedo-Soler, S.~F. Huelga, and M.~B.
  Plenio.
\newblock The role of non-equilibrium vibrational structures in electronic
  coherence and recoherence in pigment-protein complexes.
\newblock {\em Nat Phys}, 9(2):113--118, 02 2013.

\bibitem{PhysRevLett.104.207701}
Marlan~O. Scully.
\newblock Quantum photocell: Using quantum coherence to reduce radiative
  recombination and increase efficiency.
\newblock {\em Phys. Rev. Lett.}, 104:207701, May 2010.

\bibitem{kubo:174}
Ryogo Kubo.
\newblock Stochastic {L}iouville equations.
\newblock {\em Journal of Mathematical Physics}, 4(2):174--183, 1963.

\bibitem{Kubo:1962fk}
R.~Kubo.
\newblock {\em Fluctuations, Relaxation, and Resonance in Magnetic Systems}.
\newblock Plenum, New York, 1962.

\bibitem{karabunarliev:057402}
Stoyan Karabunarliev and Eric~R. Bittner.
\newblock Spin-dependent electron-hole capture kinetics in luminescent
  conjugated polymers.
\newblock {\em Physical Review Letters}, 90(5):057402, 2003.

\bibitem{Bittner:2005be}
Eric~R. Bittner, John Glenn~Santos Ramon, and Stoyan Karabunarliev.
\newblock Exciton dissociation dynamics in model donor-acceptor polymer
  heterojunctions. i. energetics and spectra.
\newblock {\em The Journal of Chemical Physics}, 122(21):214719--9, 2005.

\bibitem{karabunarliev:3988}
Stoyan Karabunarliev and Eric~R. Bittner.
\newblock Dissipative dynamics of spin-dependent electron--hole capture in
  conjugated polymers.
\newblock {\em The Journal of Chemical Physics}, 119(7):3988--3995, Aug 2003.

\bibitem{pereverzev:104906}
Andrey Pereverzev and Eric~R. Bittner.
\newblock Time-convolutionless master equation for mesoscopic electron-phonon
  systems.
\newblock {\em The Journal of Chemical Physics}, 125(10):104906, 2006.

\bibitem{tamura:021103}
Hiroyuki Tamura, Eric~R. Bittner, and Irene Burghardt.
\newblock Exciton dissociation at donor-acceptor polymer heterojunctions:
  Quantum nonadiabatic dynamics and effective-mode analysis.
\newblock {\em The Journal of Chemical Physics}, 126(2):021103, 2007.

\bibitem{tamura:107402}
Hiroyuki Tamura, John G.~S. Ramon, Eric~R. Bittner, and Irene Burghardt.
\newblock Phonon-driven ultrafast exciton dissociation at donor-acceptor
  polymer heterojunctions.
\newblock {\em Physical Review Letters}, 100(10):107402, 2008.

\bibitem{Troisi2013}
A.~Troisi.
\newblock How quasi-free holes and electrons are generated in organic
  photovoltaic interfaces.
\newblock {\em Faraday Discussions}, 2013.

\bibitem{PhysRevB.71.045203}
Xiujuan Yang, Tieneke~E. Dykstra, and Gregory~D. Scholes.
\newblock Photon-echo studies of collective absorption and dynamic localization
  of excitation in conjugated polymers and oligomers.
\newblock {\em Phys. Rev. B}, 71:045203, Jan 2005.

\bibitem{doi:10.1021/jp307040d}
Oliver~Robert Tozer and William Barford.
\newblock Exciton dynamics in disordered poly(p-phenylenevinylene). 1.
  ultrafast interconversion and dynamical localization.
\newblock {\em The Journal of Physical Chemistry A}, 116(42):10310--10318,
  2012.

\bibitem{doi:10.1021/jp307041n}
William Barford, Eric~R. Bittner, and Alec Ward.
\newblock Exciton dynamics in disordered poly(p-phenylenevinylene). 2. exciton
  diffusion.
\newblock {\em The Journal of Physical Chemistry A}, 116(42):10319--10327,
  2012.

\bibitem{bittner:034707}
Eric~R. Bittner, Stoyan Karabunarliev, and Aijun Ye.
\newblock Photoconductivity and current producing states in molecular
  semiconductors.
\newblock {\em The Journal of Chemical Physics}, 122(3):034707, 2005.

\end{thebibliography}


\end{document}